\documentstyle[11pt,paspconf]{article}

\begin{document}

\title{ICM PHYSICS AT HIGH REDSHIFT}
\author{Maguerite PIERRE}
\affil{CEA/DSM/DAPNIA/SAp Saclay,
 F-91191 Gif sur Yvette  }

\begin{abstract}
While the concept of ``clustering" primarily refers to galaxies, 
the fact is that galaxies can be neglected - at least in terms of 
mass - when considering the continuous competition between accretion
 and relaxation processes of dark matter and gas, which control the 
dynamical life of galaxy clusters. Gravity is certainly the driving
 force in cluster formation and evolution, but is not sufficient to
 explain in detail the properties of the ICM brought to light by the
 most recent X-ray data: non thermal phenomena revealed  by 
radio observations, micro-physics and feedback from galaxies play 
also a key role. From numerical simulations constrained by observations of the intra-cluster medium at low $z$, we  attempt to extrapolate properties back in the time  when density contrasts in the universe were much less pronounced. 
 
\end{abstract}


\section{Introduction: What is ICM ?}
 
In 1970, the UHURU satellite revealed that clusters are the 2nd most powerful X-ray sources in the sky, after the QSOs (Forman et al 1972). Their emission was soon recognized to be of thermal origin, produced by a very diffuse gas at a temperature of several tens of million degrees trapped in the cluster gravitational potential. During the last two decades, due to the X-ray observatories EINSTEIN (1978), GINGA (1987), ROSAT (1990) and ASCA (1993), our knowledge of the properties of this gas, commonly called ``Intra-Cluster Medium" (ICM), has made tremendous progress. The relative {\em weights} of the three cluster components  are now quite confidently estimated: \\
\underline{Galaxies} - what is seen in the optical - accounting for    $< 5\%$ of the cluster total mass, ($M_{tot}$)\\
\underline{Gas} - what is seen in the X-ray - $15\%~ M_{tot}$, with densities of the order of 1 atom/litre, temperatures $ \sim 5~10^8 K$ and abundances $\sim ~0.3$ solar\\
\underline{Dark Matter} - what is not seen - $80\%~ M_{tot}$\\
Quoted value are for a ``medium'' cluster having a mass of a few $10^{14} M_{\odot}$ (Ho = 50 km/s/Mpc); except for very small (compact) groups, the observed variations in abundances and gas mass fractions is small.
From this,  the cluster gravitional potential is defined by the distribution of the Dark Matter plus that of the X-ray gas. Amazingly, it turns out that galaxies  whose grouping originally led to the term ``cluster of galaxies" have a negligible contribution in these gravitationally bound systems.  From the theoretical point of view, clusters are described by only one parameter: they are (virialized ?) entities having masses of the order of $10^{14}-10^{15} M_\odot$ - the mass being however, not a direct observable but rather easily deduced from the X-ray temperature (Henry 1997 and references therein).  \\
Because of historical observational reasons, the ICM is generally first considered through its X-ray properties. We thus start by summarizing the main results obtained from X-ray data and the basic properties of the ICM. Section 3 presents the information provided by radio wavalengths: we shall  describe how it complements  the X-ray data in our understanding of ICM physics. Then, we briefly address a few more theoretical issues, as to the state of the ICM, which may have a decisive impact on the data interpretation.  A relevant question is also the role of galaxies: are they only test particules in the cluster gravitationnal potential? In Section 5, we  present a few simple low-$z$ correlations which appear to constrain galaxy activity and, correlatively, the state of the ICM  at high redshift.  The conclusion summarizes the current status and presents prospects for further X-ray studies of the ICM at high redshift and of very low density structures such as small galaxy groups or filaments.

\section{What X-ray PHYSICS today?}
In this Section, we recall the bases of cluster X-ray astronomy, as most of the audience at the Conference are of ``optical'' origin.  
Let's first stress that X-ray telescopes count individual photons, and so far, have  measured their energy with an accuray of the order of 50\% (ROSAT) - 20\% (ASCA) and resolved from 3 arcmin (ASCA) to 25-5 arcsec (ROSAT).
\subsection{Basic physics with $\sim 10^3$ photons}
A few thousand X-ray photons received from a galaxy cluster enable the determination  of the global properties of the ICM: surface brightness profile and some morphological information (2D), electron density profile (3D, $n_{gas}(r)$), average metallicity and temperature. 
It is also possible to estimate cluster masses under the assumptions of hydrostatical equilibrium, spherical symmetry and ideal gas:
$$   \frac{dP_{gas}}{dr} = -n_{gas}\frac{d\Phi_{grav}}{dr}$$
$$   \mu m_{H}P_{gas} = n_{gas}kT$$
$$ M_{tot}(<r) = -\frac{kTr}{\mu m_{H}P_{gas}G}\left(\frac{dlogn_{gas}}{dlogr}+\frac{dlogT}{dlogr}\right)  $$
The enclosed mass within a given radius is readily obtained from the last equation,  assuming a constant average temperature. Subsequently, two puzzling results have emerged from the observations; (1) as mentioned above, a gas fraction of the order of 15-20\% is usually found in rich clusters (Arnaud \& Evrard 1999 and references therein). If clusters are representative of the rest of the universe, this is in severe conflict with the results of the primordial nucleo-synthesis  giving $ 0.04 < \Omega_{b} < 0.06$ ... for $\Omega_{o}$ = 1 (the so-called `` baryon catastrophe'', White et al 1993); (2) there appears to be a factor 2 discrepancy between X-ray and lensing mass estimates  for some clusters, the X-ray mass being too small unless temperatures were twice those actually measured; however the agreement is good for apparently relaxed clusters (Allen 1998).

\subsection{Advanced physics with $\sim 10^{4}$ photons}
More than 10 000 collected photons enable the basic analysis described above to be applied to cluster sub-regions such sectors or concentric rings. There are however critical observational limitations to this procedure, namely: it is   far from trivial to recover the ``true'' 3D temperature and abundance profiles from the 2D projected X-ray data, having, moreover, an instrumental PSF which depends both on the photon off-axis position and energy. With the aid of shophisticated deconvolution procedures, ROSAT and ASCA data have provided interesting information about the radial variation of temperature (such as the existence of a possible universal temperature profile) and abundances in a series of nearby clusters (Markevich et al 1998).\\
Some, apparently  relaxed, clusters show a strong X-ray peak at their very center associated with a cooler temperature, a phenomenon commonly known as ``cooling flow'' (CF) (see Fabian 1994 for a review): Once a cluster size object has collapsed, the hot gas tends to reach   hydrostatic  equilibrium. But it radiates energy and cools (the cooling coefficient, $\Gamma_{X}$, or X-ray emissivity is proportional to the density square). The more it radiates, the more it contracts and the more its density increases. This has the immediate effect of most stronlgy enhancing the cluster central density, thus its X-ray  brightness, and, ultimately, would lead to the so-called ``cooling catastrophe''. Because of the pressure gradient which develops, the gas is assumed to flow inwards, in order to stay isobaric. The fate of the accumulating gas at the center is however an enigma, as star formation rates usually found to be associated with the cluster dominant galaxy, through optical emission lines and blue nebulae, are 10 to 100 times lower than inferred from the X-ray mass flow rate estimates. ASCA spectral observations of CF clusters need at least two temperatures to be correctly fitted and reveal the presence of a central Hydrogen absorption (Allen et al 1996), hence suggesting that the ICM is actually multi-phase and could cool very low. We shall come back to this question in Sec. 4.

\subsection{What is HIGH REDSHIFT for '99 cluster X-ray astronomy?}
To summarize the Introduction,  presently - i.e. with ASCA and ROSAT - {\em basic} physics is achievable out to $ z \sim 0.5$ whereas {\em advanced} physics is performed out to $ z \sim 0.1$. Note, also, that it is currently possible to detect clusters out to $ z \sim 1$ (with a few tens of photons).   \\
The validity of the hydrostatic equilibrium assumption is certainly questionable, but with the present data, hard to test. The comparison with lensing mass estimates - free of this hypothesis - suggests that this may not be always verified. We shall see in the next sections that the physics of the ICM is actually much more complex than suggested from the X-ray viewpoint alone.

\section{Radio observations and magnetic field}
\subsection{Individual synchrotron radio sources}
Radio galaxies produce relativistic particles, that interact with the ICM. Indeed, head-tail (HT) and wide-angle-tail (WAT) galaxies are exclusively found in cluster environments: While the straight shape of the HT galaxies (usually located at the cluster periphery $\rightarrow$ large peculiar velocity) is explained by dynamical frictions against the ICM, the origin of the bending of the WAT (always cluster dominant galaxy $\rightarrow $ almost at rest in the cluster frame) is still unclear and one has to invoke buoyancy forces or winds in the ICM.\\ 
The combined examination of high resolution X-ray and radio maps reveals how complex the mutual interactions can be. For instance, the radio morphology of the central galaxy of the Perseus cooling flow cluster, NGC 1275, shows a clear anticorrelation with the surrounding X-ray gas emissivity, hence indicating that the relativistic particles repell the hot gas (B\"ohringer et al 1993). To the contrary, the central galaxy of Virgo, M87, exhibits a similarly obvious correlation; here the excess of X-ray thermal emission could be produced by a cool gas component due to the combined action of the radio jets and the magnetic field (B\"ohringer et al 1995). Note that recent 90 cm observations of M87 show a very extended and bubble-like structure, suggesting the presence of an ``outflow'' (Owen et al 1999).  \\
To quantify the interactions of the radio lobes with the ICM, one way is to test the equipartition hypothesis, which is satisfied if the energy density of the relativistic particles equals that of the magnetic field or, in other words: when the minimum pressure deduced from the radio data equals the X-ray pressure. Although the estimate of the minimum radio pressure is subject to a series of hypotheses - certainly more ad hoc than for the X-ray pressure derivation - there appears to be a large variety of situations (Liang et al 1997  and references therein), as suggested by the above morphological considerations. 

\subsection{Diffuse cluster sources}
In addition to cluster radio galaxies, extended diffuse radio emission is observed on cluster scales. These so-called halo or relic sources are rare events; only some 20 cases are known. They all have a steep radio spectrum and are found in merging or post-merging clusters (for a review see Ferreti et al 1995). The nearest example is the Coma cluster (Ferreti 1999), one of the most remarkable and well studied case is found in A2256 (Röttgering et 1994; for the X-ray temperature map: Briel \& Henry 1994), the most distant one being in A1300 ($z = 0.3$, Reid et al 1999). A possible explanation for the origin of halos would be the acceleration of cosmic rays, contained in the ICM, at the merging front. Indeed, only a few hundredths of the shock power would be enough to produce the energy of the electron population of the radio halo (B\"ohringer et al 1992). \\
The comparison of the X-ray and radio maps of A85 revealed an X-ray bump coinciding with an extended Very Steep Spectrum Radio Source not associated with a galaxy. Here, the diffuse X-ray component is likely not to be thermal, but produced by inverse Compton effect:  the primeval 3K CMB photons, scattering off the relativistic electrons of the diffuse radio source can produce the X-ray emission at the observed level (Bagchi et al 1998).

\subsection{Cooling flows}
HST/WFPC2 observations of the CF cluster, A1795, revealed a chain of blue knots coinciding with the radio lobes of the cD galaxy (Pinkey et al 1996). A simple interpretation suggests that star formation is triggered by the radio activity in clumps where the multi-phase ICM is cooler and denser. As mentioned earlier, cooling flows are thought to be considerable sources of cold gas, which can be detected in the optical and X-ray either in emission or in absorption; moreover, the dense inflowing gas may also feed an AGN. Assuming that the production of relativistic electrons by the central engine is related to the gas accretion rate, one can imagine a feedback mechanism inhibiting the accreation, thus, leading to a cyclic regime with phases of accretion where the radio activity is turned-off and phases of production of particles and no accretion; this would explain the observed fraction of CF clusters possessing a central radio source ($\sim 70\%$) and why so little star formation is usually observed in the optical (Tucker \& David 1997). 

\subsection{$^{57}Fe$ Hyperfine transition}
Sunyaev \& Churazov (1984) have predicted the existence of a hyperfine transition for the lithium-type ion  $^{57}Fe^{+23}$ at $\sim$ 0.3 cm. Its spontaneous emission coefficient is significanlty higher than that of HI and thus would partly compensate the low abundance of the $^{57}Fe$ isotope. The detection of this line would be most interesting for our knowledge of the ICM: direct insight into the dynamics and turbulent motions, alternative estimate of the metal radial abundance, accurate and straightforward measurement of the cluster redshift. However, the predicted intensity of the line is well below 1 mK for a strong CF cluster like A85 and, so far, the line has not been detected (Liang et al 1997).  

\subsection{Magnetic field}
Magnetic fields are estimated via Faraday rotation, mostly in very nearby clusters to ensure  the presence of enough background radio sources. The strength of the field is of the order of a few $\mu$G with a scale length of about 10pc (Kim et al 1991). Strong rotation measures are found in CF clusters. Faraday rotation and gamma observations have led Ensslin et al (1997) to show that the cluster radio galaxies could sustain the ICM with injection of cosmic rays and magnetic field. Non-thermal pressure could then affect the simplistic cluster mass calculation in X-ray, requiring a higher total mass, and thus reduce the discrepancy between X-ray and lensing estimates; it would also tune down the ``baryon catastrophe'' (cf Sec. 2.1).   

\subsection{The Sunyaev Zel'dovich effect}
The Sunyaev Zel'dovich (S-Z) effect is due to Inverse Compton scattering of the CMB photons on the ICM electrons and, thus, produces a distortion of the CMB blackbody spectrum, when observed through a cluster. The amplitude of the thermal effect measured as a CMB temperature variation, is:
$$ y = \int \sigma_{T}n_{e}\frac{kT_{e}}{m_{e}c^{2}}dl$$
Where $n_{e},T_{e}, l$ are respectively the ICM electron density and temperature, the pathlength along the line of sight through the cluster, and $\sigma_{T}$, the Thomson cross-section. The $y$ parameter is thus representative of the cluster electron pressure. The effect is independent of the cluster distance and despite its weakness ($\Delta T_{MBG}/T_{MBG} \sim 10^{-4.}$) it is now currently possible to map clusters with radio interferometers (Carlstrom 1999).  There is a recent review by Birkinshaw (1999) on the topic, so we shall just mention a few points here, in direct connection with the ICM Physics.\\ 
The S-Z decrement (which  is independent of distance), combined with the knowledge of the X-ray temperature and electron density radial profiles  (distance dependent) was proposed as a straightforward  way to measure $H_o$. Actually, there are numerous limiting factors in the use of this method, some of them being of technical nature, others directly related to the ICM complexity.\\
-1- The S-Z effect differs significantly from the X-ray data in their derived properties of the cluster atmosphere: 
$S-Z \propto n_eT_e$ while  $\Gamma_X \propto n^{2}_{e}T^{1/2}_e $. They are thus likely to have different angular structures, which should provide information on the local variations of temperature and density in the cluster gas. Improvements in X-ray technology, which provides spatially resolved spectroscopy, have largely superceded S-Z information. However, the S-Z drops off less rapidly than the X-ray surface brightness ($S_X$) and thus could still be an important probe of structure in the outer parts of clusters; but the current sensitivity of S-Z instruments is too low relative to the sensitivity of X-ray image and spectra, for this to be useful in most cases.\\
-2- Where the cluster contains a radio source (particularly a radio halo) the S-Z effect is of particular interest since it provides a direct measurement of the electron pressure near that radio source and so can be used to test whether the dynamics of the radio emitting plasma are strongly affected by the external gas pressure.\\
-3- On the smallest scales (where ICM structures are unresolved by X-ray or radio telescopes) the structures are better described by a clumping of the gas and the S-Z effect and $S_X$  scale differently. For instance, if clumping is isobaric, with the pressure in clumps the same as outside, then the S-Z effect will show no variations in regions where the gas is strongly clumped while  $S_X$ will increase as $n_e^{3/2}$. No useful results on the clumping of cluster gas have been reported in the literature to date. \\
-4- One direct use of the S-Z effect is as  probe of the gas mass enclosed within the telescope beam. The surface mass density in gas can be related to the $y$ parameter if $T_e$ is constant. This can be compared directly with the mass estimates derived from the study of gravitational shear in clusters.

\section{Some theoretical considerations}
Having presented the main observational properties of the ICM, we show here by two examples, how current results and views need to be sharpened by theoretical considerations. 
\subsection{The state of the ICM}
The thermodynamical state of the ICM is of primary importance in the study of the formation and evolution of cosmic structures, especially at high redshift. As in the collapse, the gas velocity becomes supersonic, shock fronts form at about the virial radius and separate the infalling from the inner gas already at viriral temperature. 
In the dense central region, cooling is likely to play a dominant role. In the outer regions, the density is rather low and allows an adiabatic treatment of the gas dynamics and, at the same time, non equilibrium thermodynamics occurs in this hot and diffuse plasma. Takizawa (1998) and Chièze et al (1998)  have shown that whereas the hydrostatic and thermodynamical equilibrium assumptions are valid in the central regions, they are both violated in the outer regions: A significant decoupling between electrons and ions occurs between $r_{200}/2$ and the shock front\footnote{$r_{200}$ is the radius where the density equals 200 times the critical density}, located roughly at 2$r_{200}$. The maximum departure is located at $r_{200}$ and is about 20\%. Therefore, the usual assumption of thermodynamical equilibrium between electrons and ions breaks down in this region. This could lead to a thermodynamical decoupling of 50\% which means that the observed electron temperature (measured in the X-ray) underestimates by a factor of 2 the true dynamical temperature. Therefore, the error in cluster mass estimates due to a departure from thermodynamical equilibrium could be as large as a factor of 2.
\subsection{The multi-phase medium}
In Sec. 2.2 we mentioned observational evidences for a ``several-temperature'' medium in cooling flow clusters. Actually, Nulsen as early as 1986 predicted the existence of a multi-phase medium on cluster scales. A collapsed object radiates energy, but the cooling process is highly unstable and leads to the formation of cool and dense blobs and filaments, evolving from an initial temperature fluctuation distribution. As cooling proceeds, a net mass flow from the hot phase to the cool phase develops, although some hot, diffuse gas always remains in the cluster. This remaining X-ray emitting gas inhibits the ``cooling catastrophe'' at the cluster center by increasing entropy  and creating a core. After a few cooling times, a self-similar evolution may build up, allowing  the description of a multi-phase cooling flow using very simple analytical models (Teyssier 1996). Cooling processes will be much more efficient (dominating) in groups. The remaining hot and diffuse gas has an X-ray luminosity significantly reduced compared to the equivalent ``adiabatic" cluster or group. The X-ray mass fraction is also strongly influenced by this multi-phase evolution.

\section{A glimpse at high-redshift conditions}
A few basic correlations between cluster properties ($L_{X}, ~T_{X},~ \sigma_{v}$...), have been investigated out to $z \sim 0.4$.  There does not appear to a conspicuous evolution, either in the correlations (e.g. Mushotzky \& Scharf 1997) or in the cluster luminosity function alone (e.g Rosati et al 1998, out to $z \sim 0.8$). However these local correlations described below provide clues about the state of the ICM in the era of cluster formation.\\
\underline{The $L_{Bol}-T$ relationship}\\
A clear correlation between temperatures and X-ray luminosities exists for X-ray clusters. The scatter is large, but it is remarkable that the correlation extends from very luminous clusters down to groups, thus over the luminosity range $ 5~10^{45}~-~5~10^{42}$ erg/s,  following $L_{Bol} \propto T_{X}^{3.4}$ (David et al 1993). Arnaud \& Evrard (1999) have shown that restricting to non CF clusters and to objects having well measured temperature, the correlation is very tight, with a slope $\sim 3$. However, this average relationship is rather far from that expected for a simple gavitational collapse, namely: $L_{Bol} \propto T_{X}^{2}$. Non gravitational ``preheating'' processes, such as shocks or feedback from supernova explosions at early stages, have been proposed in order to raise the temperature of the groups before they merge into clusters (Kaiser 1991). The energy contribution due to winds will be larger in cool clusters (groups) than in hot ones, thus, producing the observed steepening of the $L_{Bol}-T$ relationship in the local universe.
More quantitatively, hydrodynamical simulations show that including feedback raises the entropy of the ICM, preventing it from collapsing to densities as high as those obtained in the pure infall model (Metzler \& Evrard 1994). The effect is most pronounced in subclusters formed at high redshift, clusters with feedback being always less luminous but, then, they experience a more rapid luminosity evolution. Also, the simulations suggest the presence of a radial iron abundance gradient as a consequence of ejection which takes place while galaxies have a gradient in number density with respect to the primordial gas (as the ICM is slightly hotter and therefore more extended than the galaxy distribution.)\\ 
 \underline{The $\sigma_{v}-T$  relationship}\\
Assuming that galaxy orbits are isotropic, that the gas and galaxies occupy the same potential well and that gravity is the only source of energy,  we can predict that the galaxy velocity dispersion ($\sigma_{v}$) and the temperature of the ICM should be correlated by $\sigma_{v} \propto T^{0.5}$. It is found that $\sigma_{v} \propto T^{0.61~or ~0.76}$ (Bird et al 1995).
Simulations show that both energetic winds and galaxy velocity  bias can result in the observed $\sigma_{v}-T$  relation.\\
\underline{The $R_{isophot}-T$ relationship}\\
From a sample of ROSAT PSPC clusters, Mohr \& Evrard (1997) have shown that there is a tight relation between some measure of the cluster X-ray isophotal size and its temperature. This indicates that the ICM structure outside the core regions is a well-behaved function of the temperature, with no evidence that CF significantly affect the relationship, and suggests a tight correlation betweeen the temperature and the virial mass. However, the slope of the $R_{isophot}-T$ relationship is steeper than the slope inferred from numerical simulations. Again here, introducing galaxy feedback produces the kind of ICM structural changes required to steepen the correlation.\\

\noindent
Statistical and evolutionary properties of cluster samples are potentially powerful tools to constrain cosmology (Blanchard, these proceedings). Conversely, the three above-mentioned local correlations (all involving the temperature parameter) indicate that statistically complete surveys extending to $z \sim 0.5-1$ would provide  constraining power on the amount of feedback in clusters, although in a cosmologically dependent fashion, which still needs to be understood.

\section{Conclusion}
We have presented the main aspects of the current knowledge about ICM and its physics, primarily from the X-ray point of view. Hydrostatic equilibrium seems a reasonable working hypothesis within the (inner) regions of ``regular'' clusters explored so far... at least if one is statisfied with an uncertainty of a factor of two, which may be considered  excellent for astrophysical accuracy standards or may point out some degree of over-simplification. Indeeed, information provided by the radio wavelengths complements and refines this view, suggesting that phenomena such as turbulence and magnetic field may have  a non negligible role; Sunyaev Zel'dovich effect studies are potentially a powerful complement to X-ray data.\\ 
 In addition, recent cluster observations in the Extreme UV (Sarazin \& Lieu 1998) and Hard X-ray (Fusco-Femanio et al 1999), show emission in excess to what is expected from a purely thermal spectrum extrapolated from the X-ray band. The excesses would be due - respectively - to Inverse Compton diffusion of low energy cosmic ray present in the ICM and to thermal bremsstrahlung produced by an electron population having energies above 50 keV, accelerated by turbulence.  \\  
Despite their negligible mass, galaxies appear to play an important role in the physics of the ICM, especially at high redshift. (i) Feedback  from star forming activity preheats the ICM via winds and supernova explosions prior to cluster formation;  the effect of individual galaxies  is stronger in groups than in clusters and remains conspicuous in the low $z$ group population (Davis et al 1999) (ii) metals found in the ICM originate in galaxies: due to ASCA's good energy resolution enabling the measurement of the Oxygen abundance, it has been shown that type II supernovae are responsible for a large part for the enrichment of the ICM (Loewnestein \&  Muschotzky 1996); (iii) radio galaxies poduce relativistic particles that interact strongly with the ICM. (iv) the role of the central (radio) galaxy appears to be crucial in CF clusters. To conclude, we emphasize that the physical state of the ICM is inexorably linked with galaxy formation hence, with star formation.\\
 
Investigating ICM X-ray physics at high redshift  will require large collecting areas and good energy resolution. This is what is to be offered by XMM, the 2nd ESA Corner Stone.  The {\em basic} physics mentioned in Sect. 2 will be possible out to $ z \sim 1$, while {\em advanced} temperature mapping will be achievable for the more nearby clusters out to the ``virial radius'' where complicated phenomena are supposed to take place (Sec. 4). With moderate exposure times it will be possible to detect the entire cluster/group population out to  $ z \sim 0.5$ and study in detail the effect of the feedback and cooling processes (Pierre et al 1999). Correlativeley, the very high spatial and spectral resolution of Chandra should enable the study of fine morphological ICM details in nearby objects as well as possible soft X-ray absorption lines (in QSO spectra) from warm/cold gas present in clusters or cosmic filaments (Cen \& Ostriker 1999).  

\acknowledgments
It is a pleasure to thank Romain Teyssier for many cinteresting discussions and S. Madden for useful comments.

\end{document}